\documentclass[twocolumn,10pt,superscriptaddress,pra,showpacs]{revtex4}
\usepackage{amssymb,amsmath,graphicx,epsfig,subfigure}
\usepackage{times}
\begin{document}

\title{Anomalous decoherence effects in driven coupled quantum spin systems}

\author{Chuan-Jia Shan}
\affiliation{College of Physics and Electronic Science, Hubei Normal
University, Huangshi 435002, China}
\author{Pan-Pan Wu}
\affiliation{College of Physics and Electronic Science, Hubei Normal
University, Huangshi 435002, China}
\author{Wei-Wen Cheng}
\affiliation{Institute of Signal Processing and Transmission, Nanjing University of Posts and Telecommunication, Nanjing 210003, China}
\affiliation{College of Physics and Electronic Science, Hubei Normal
University, Huangshi 435002, China}
\author{Ji-Bing Liu}
\affiliation{College of Physics and Electronic Science, Hubei Normal
University, Huangshi 435002, China}
\author{Tang-Kun Liu}
\affiliation{College of Physics and Electronic Science, Hubei Normal
University, Huangshi 435002, China}
%\maketitle
\date{\today}

\begin{abstract}
We discuss anomalous decoherence effects at zero and finite temperatures in driven coupled quantum spin systems. By numerical simulations of the
quantum master equation, it is found that the entanglement of two coupled spin qubits exhibits a non-monotonic behaviour as a function of the noise strength. The effects of noise strength, the detuning and finite temperature of independent environments on the steady state entanglement are addressed in detail. Pumped by an external field drive, non-trivial steady states can be found, the steady state entanglement
increases monotonically up to a maximum at certain optimal noise strength and decreases steadily for higher values. Furthermore, increasing the detuning can not only induce but also suppress steady state entanglement, which depends on the value of noise strength. At last, we delimit the border between presence or absence of steady state entanglement and discuss the related  experimental temperatures where typical biomolecular systems exhibit long-lived coherences and quantum entanglement in photosynthetic light-harvesting complexes.
\end{abstract}

\maketitle
\section{Introduction.}

Decoherence, induced by the coupling between a quantum qubit and its surrounding environment, is a main obstacle to the practical realization of quantum information processing\cite{Nielsen}. The controlled generation and detection of entanglement of quantum states remains one of the fundamental challenges of quantum physics. It is of great importance to analyze the entanglement decay induced by the unavoidable interaction with the environment. In recent years, there have been many investigations of decoherence\cite{Yu,experimental,Ficek,Lopez,Xu,Mazzola,Dijkstra,Jing,Shin}, careful investigation of well-understood model systems continue to produce surprises that add to fundamental understanding. For example, Entanglement sudden death (ESD)\cite{Yu} have been addressed in different quantum systems. Moreover, it has also been experimentally observed \cite{experimental}. Entanglement sudden birth (ESB) \cite{Ficek,Lopez} is the creation of entanglement where the initially unentangled qubits can be entangled after a finite evolution time. Traditionally, it has been assumed that noise can only have detrimental effects in quantum information processing. Recently, however, it has been suggested, and realized experimentally, that the environment can be used as a resource \cite{Huelga,Verstraete,Kraus,Plenio,Wolf,Bellomo,Hartmann}. In particular, it is generally believed that stronger noise causes severer decoherence. Strikingly, recent theoretical results \cite{Liu,An} suggest that under certain conditions, the opposite (an anomalous decoherence effect) is true for spins in quantum baths. Furthermore, an experimental observation of an anomalous decoherence effect for the electron spin-1 of a nitrogen-vacancy centre in high-purity diamond at room temperature has been reported \cite{Du}. This discovery establishes the controllability of quantum baths and paves the way for exploiting spin ensembles in quantum information processing. As a result, there has been an increasing interest in better understanding the interplay between coherent and incoherent quantum dynamics arising from environmental interaction.

The entanglement dynamics of open quantum systems may be rather complex, mostly due to the structure of the environment interacting with the quantum system. In the particular case of the dissipation and decoherence phenomena, the Lindblad or Bloch-Redfield master equations \cite{Breuer} have been used as the common approaches to study the effects of the environment on the entanglement dynamics.  Generally, the non-unitary evolution of the reduced-density matrix of the system can be obtained after tracing out the environmental degrees of freedom. In this process, some approximations (the weak coupling and Born-Markov approximations) are often made in the derivation of a master equation.  Recent reservoir engineering techniques aim to alter the dynamics of dissipation and decoherence in an open quantum system \cite{Poyatos,Diehl,Prior,Wang,Wu}. The coupling of the quantum system with its surrounding environment and the induced entanglement decay motivate some important questions, such as how to understand its sources and possibly how to find ways to circumvent it. Therefore, it is of fundamental and practical importance to study the decoherence dynamics of a system in structured reservoirs.

In recent years, quantum theories have been developed to treat the decoherence problem in a mesoscopic quantum bath. These quantum theories have been
successful in studying decoherence in various systems and predicted some surprising quantum effects \cite{Yao,Huelga and Plenio,Liao}.  A number of researches have indicated the quantum nature of nuclear spin baths in the presence of a classical driving field. Some important theoretical and experimental works dealing with the effects of the driving on the coherent dynamics have recently appeared \cite{Lambert,Li,Galve,Cai}.  In this paper, we discuss the entanglement dynamical behavior of two driven coupled qubits via a Heisenberg Ising interaction, which are connected with two independent finite temperature heat baths. The main purpose and motivation of the present letter is try to answer the following question: how entangled steady states of dissipative qubits can be generated  by adding a classical driving field to the system. In a driven dissipative system, i.e. pumped by an external coherent drive, non-trivial steady states can be found. By numerically solving the master equation, we show here that two critical noise strength $\Gamma_{c}$  and $\Gamma_{m}$ exist and noise dissipation can also have exactly the opposite effect: it can  be used to engineer a large variety of strongly correlated states in the steady state, which suggests that it is advantageous to maintain a finite, not necessarily minimal, noise strength to observe stationary entanglement experimentally. The coherent drive affords great flexibility in generating entangled states since it provides freedom in choosing the detuning and strength of the field. Without the laser field, entanglement of dissipative qubits will be destroyed but with the addition of the laser field, certain optimal noise strength and the detuning lead to the high stationary entanglement. These results enlarge the domain where stationary entanglement can exist and should be observable  at reasonable experimental temperatures.  The aim is to engineer those noise strengths and finite temperature, so that the environments drive the system to a desired final state after some time without having to actively control the system. Ultimately, this understanding may facilitate the development of finite temperature noise-assisted devices and, potentially,  quantum coherence and entanglement in light-harvesting systems.

\section{Hamiltonian of the model and entanglement negativity.}
\indent   We consider this model consists of an array of $N$ driven, coupled spin-1/2 qubits. The system is
subject to a noisy environment modeled by an infinite collection of harmonic oscillators described by creation and annihilation operators
$(a^{i}_{k})^{\dag}$  and $a^{i}_{k}$  with frequency $\omega^{i}_{k}$. This situation leads to decoherence for all degrees of freedom, unlike the common bath case. Besides, we avoid extra correlations between qubits induced by the common bath. The global Hamiltonian is written as ($\hbar=1$),
\begin{eqnarray}
H &=& -\sum_{i=1}^N \frac{\omega_0^i}{2} \, \sigma_z^i + \sum_{k,i}
\omega_k^i (a_{k}^i)^{\dagger} a_k^i + \sum_{i=1}^N \sigma_x^i  X^i
 \nonumber \\ &-& \sum_{i=1}^{N-1} J \sigma_z^i \otimes
\sigma_z^{i+1}+ \sum_{i=1}^N  \Omega_i (\sigma_{+}^i e^{-i
\omega_L^i t} +h.c.),
\end{eqnarray}
where $J$ is the coupling constant in z component of the nearest neighbor qubits, $\sigma_{x}^{i},\sigma_{y}^{i},\sigma_{z}^{i}$  are the
Pauli operators at the $i$-th qubit, $\sigma_{+}^i=|1\rangle_{i}\langle0|$. The real coefficients of the direct coupling between the two qubits are tunable parameters and can be implemented with trapped ions chains or cold atoms in an optical lattice \cite{Porras}.  $X^i=\sum_k g_k (a_k^i +a_k^{i \,\dagger})$ denotes the bath's {\em force operator}. The external driving is parameterized by its intensity, as given by the Rabi frequency $\Omega_i$, and the detuning from the qubit frequency $\delta_i=\omega_L^i-\omega_0^i$. We will consider situations where the driving is weak and the external Rabi frequency is smaller than the interqubit coupling, $\Omega < J$. Within the rotating wave approximation, weak system reservoir coupling and Born-Markov approximation, we obtain an effective Hamiltonian for the $N$-qubit array in the interaction picture,
\begin{eqnarray}
H_{eff}=\sum_{i=1}^N \frac{\delta_i}{2} \, \sigma_z^i-
\sum_{i=1}^{N-1} J \, \sigma_z^i \otimes \sigma_z^{i+1}+\sum_{i=1}^N
\Omega_i \, \sigma_{x}^i\nonumber \\-i\sum_{i=1}^N \Gamma_i
(\bar{n}+1)\, \sigma_{+}^i \sigma_{-}^i -i\sum_{i=1}^N \Gamma_i \,
\bar{n} \,\sigma_{-}^i \sigma_{+}^i
\end{eqnarray}
the quantum master equation of time evolution reads:
\begin{eqnarray}
\dot{\rho}=-iH_{\rm eff}\rho + i\rho H^{\dag}_{\rm eff} &+& \sum_i
2\Gamma_i (\bar{n}+1)\sigma_{-}^i \rho\sigma_{+}^i \nonumber \\ &+&\sum_i
2\Gamma_i \bar{n}\sigma_{+}^i \rho \sigma_{-}^i,
\end{eqnarray}
The noise strength on qubit $i$ at a temperature $T$ is given by the product $\Gamma_i \, \bar{n}$, where the explicit functional form of
the decay rate $\Gamma_i$ depends on the spectral properties of the bath and $\bar{n}$ denotes an effective {\em boson} number that depends on the bath's temperature $T$; both parameters are, in principle, controllable. This master equation treatment is valid in the parameter regime
$\Omega_i/\omega \ll 1, \Gamma_i \bar{n}/\omega \ll 1, \delta_i/\omega \ll 1$ and $J/\omega \ll 1$, where $\omega=\min\{\omega_0^i, \omega_c\}$ for a suitable frequency cut off $\omega_c$ of the bath, and all expression in this paper will then be correct to lowest non-trivial order in $\Omega_i, \Gamma_i, \delta_i$ and $J$ \cite{cohen}.

Since decoherence process leads the pure quantum system state to mixed states, in order to discuss the entanglement dynamics and steady-state entanglement in the above system, we use the negativity  as a popular measure of bipartite entanglement. The negativity under partial transposition of a two-qubit state $\rho$ is defined as \cite{Vidal}
\begin{equation}
E(\rho)=\frac{||\rho^{T_{A}}||-1}{2}
\end{equation}
This measure is based on the trace norm of the partial transposition $||\rho^{T_{A}}||$ of the state $\rho(t)$. From the Peres-Horodecki criterion of separability \cite{Peres,Horodecki}, it notes that if $\rho^{T_{A}}$ is not positive, then the state $\rho(t)$ is entangled. The negativity is an
entanglement monotone and equivalent to the absolute value of the sum of the negative eigenvalues, i.e. $E(\rho)=- 2 \sum \lambda_{i}$, where $\lambda_{i}$ are the negative eigenvalues of  $\rho^{T_{A}}$.

\begin{figure*}
\includegraphics[width=.5\linewidth]{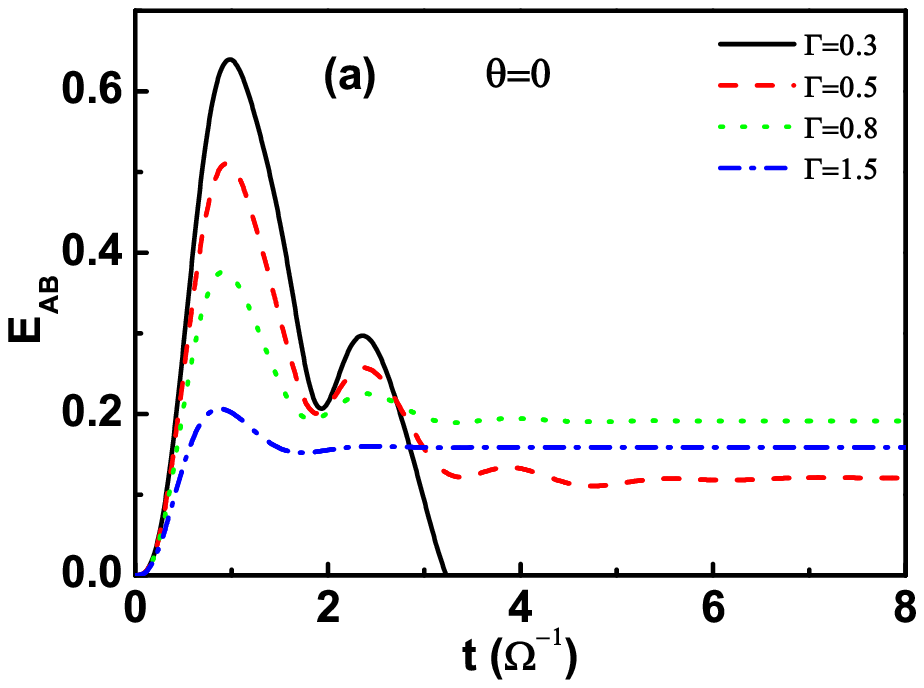}\includegraphics[width=.5\linewidth]{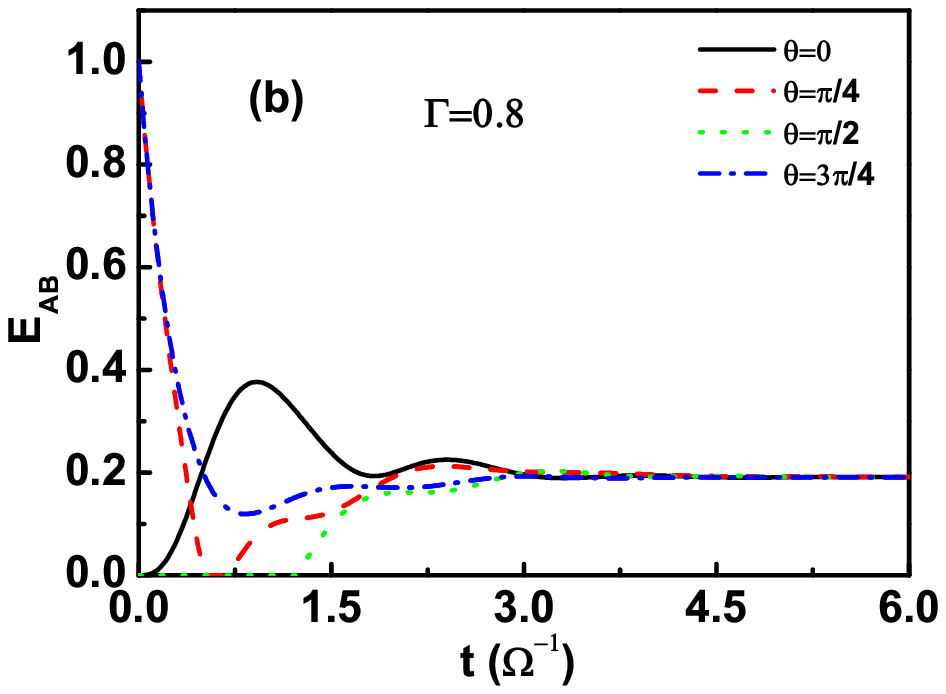}
\includegraphics[width=.5\linewidth]{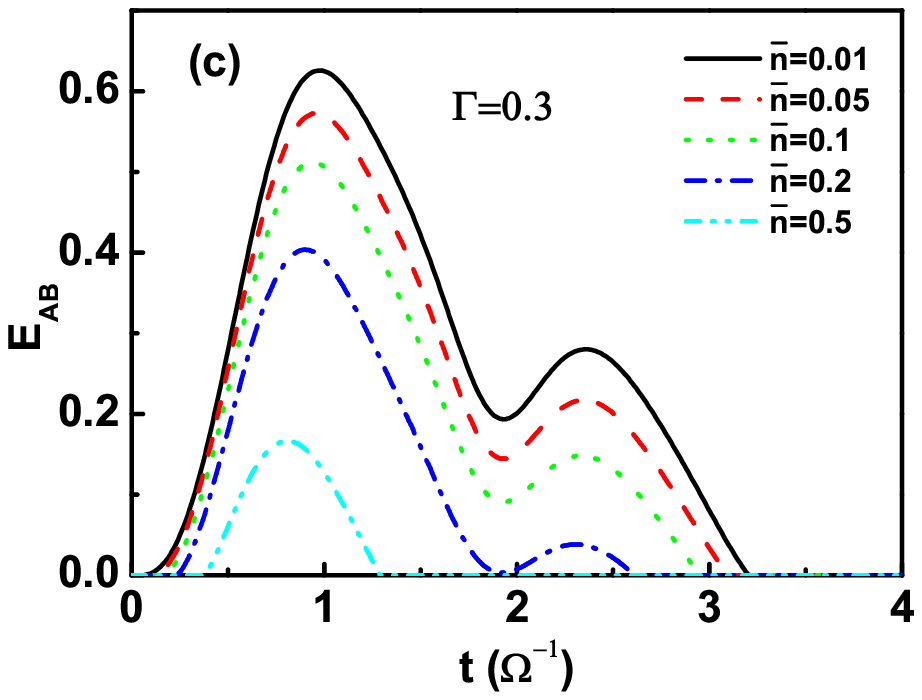}\includegraphics[width=.5\linewidth]{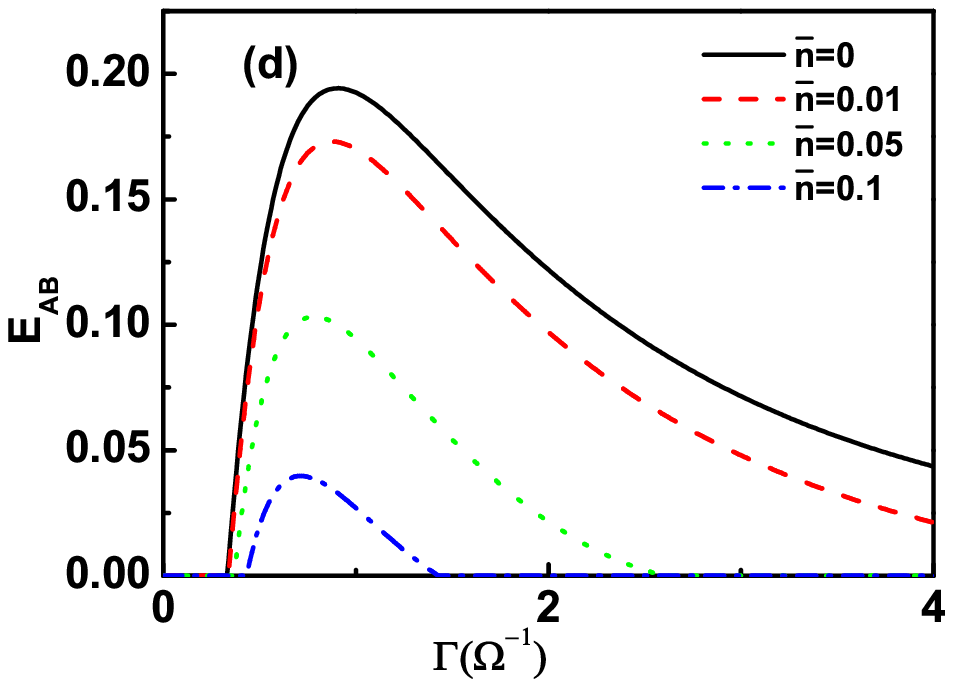}
\caption{Entanglement time evolution for two weakly driven qubits
with the coupling of strength $J/\Omega=1.5$ and the detuning $\delta_i=0$. (a),(b): $\bar{n}=0$. (c): at different mean thermal boson number $\bar{n}$. (d): the bipartite entanglement in the steady state as quantified by the $E_{AB}$ as a function of the noise strength $\Gamma$. If the noise strength $\Gamma$ is sufficiently large, the system is
inseparable in the steady state.}
\end{figure*}

\section{Results and discussions }
For this case, it is difficult to give the analytical solutions about the master equation. So, here we solve numerically the master equation with different parameters and give out the entanglement properties. Let us consider the first case where $N=2$, $\delta_{i}=0$ and $\bar{n}=0$ ($T=0$). The negativity $E_{AB}$ of two qubits as a function of time is plotted in Fig.1 for fixed values of the coupling $J$ and the driving $\Omega=\Omega_1=\Omega_2$, from which we can see the following interesting results.
Firstly, it is found that the system of weakly driven qubits, initially prepared in their ground state, develops quantum entanglement in time. In Fig.1(a), we can see that the two-qubit state can evolve into a stationary entangled state under the noise strength from initial
unentangled state. In other words, decoherence drives the qubits into a stationary entangled state instead of completely destroying
the entanglement. Moreover, we observe that the system will be entangled in the steady state only for certain finite values of $\Gamma$. Perhaps surprisingly, it is the larger value of the noise strength that yields steady-state entanglement. With the increasing of the coupling constant, the decay is suppressed, which is dramatically different from one's expectation that a stronger coupling always induces a severer decoherence. The system is entangled, and have a negative partial transpose, only if $\Gamma>\Gamma_{c}$, where $\Gamma_{c}$ is the noise threshold. If $\Gamma<\Gamma_{c}$, the state is separable. This behaviour is also illustrated in Fig.1(d) where the black solid line corresponds to the bipartite entanglement in the steady state as quantified by the $E_{AB}$ as a function of the noise strength $\Gamma$. As a result of the constraint, any entanglement measure exhibits an initial domain of vanishing entanglement for weak noise where the state is separable.  When $\Gamma$ rises above threshold, the steady state entanglement increases monotonically up to a maximum at certain optimal noise strength $\Gamma_{m}$ and decreases steadily for higher values of $\Gamma$. Therefore, this can provide us a feasible way to manipulate and control the entanglement by changing the external noise strength. we present a microscopic explanation for as a physical insight of the counterintuitive phenomenon. The origin of the stationary entanglement can be traced back to the structure of the eigenstates of the effective Hamiltonian. When the values of both reservoir noise strength $\Gamma$ are low, because of the external driving, the final state is the equally weighted superposition states, which loses coherence, so the entanglement decreases to zero. Secondly, for various values of the parameter $\theta$, corresponding to the two qubits are initially in different entangled states, we find that under some initial conditions, the entanglement of two qubits can fall abruptly to zero, and will recover after a period of time. Therefore, the ESD appears and is related to  the initial state. Even though the initial system has the same entanglement, the evolution is also different.  When the two qubits are initially prepared in their excited state, i.e. $\theta=\frac{\pi}{2}$ , the result is quite different. The entanglement versus parameter $t$ is plotted in Fig.1(b), indicating a threshold value of parameter $t$, only above which  entanglement negativity begins to be nonzero, i.e. the quantum correlation starts to appear. This is the example of phenomenon of delayed sudden birth of entanglement.  Lastly, despite the presence of decoherence, the results in Fig. 1(b) show that the entanglement  reaches the same steady value, after some oscillatory behavior, for a given set of system parameters regardless of the initial state of the system.
\begin{figure}
\includegraphics[width=\linewidth]{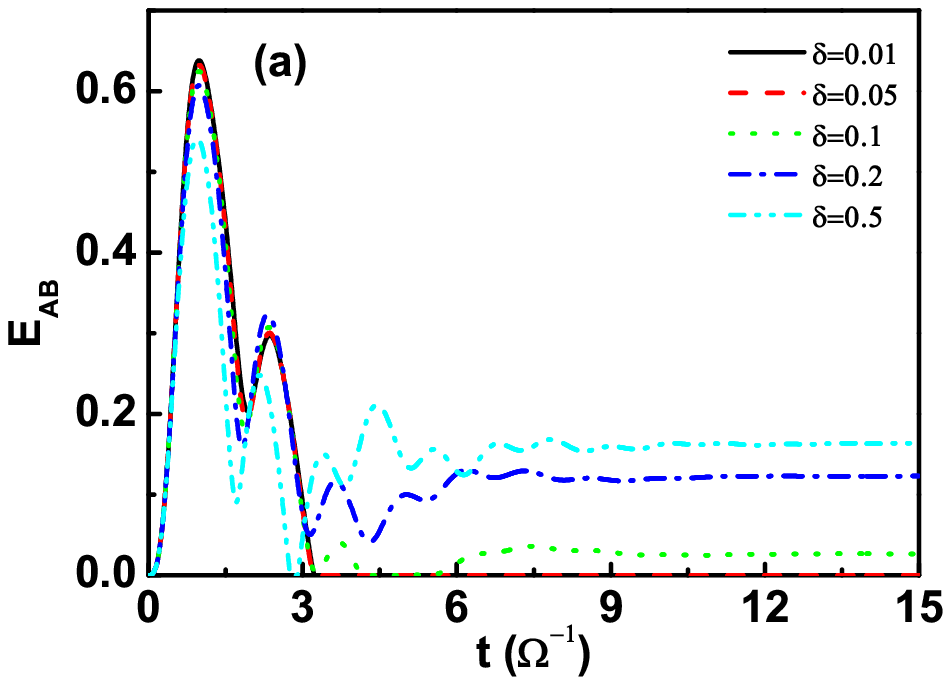}
\includegraphics[width=\linewidth]{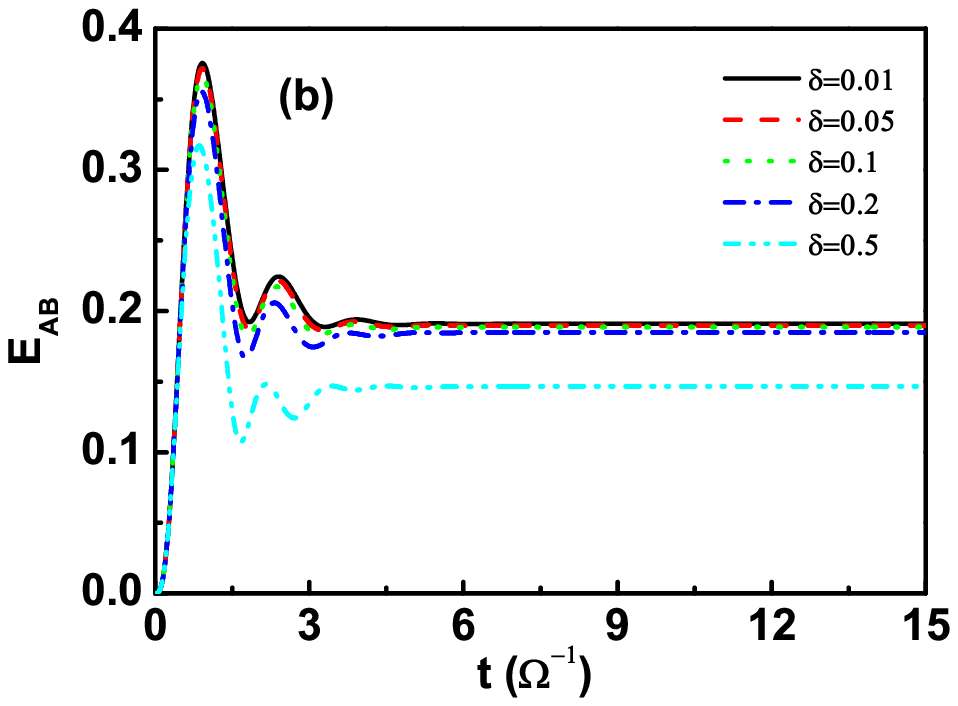}
\includegraphics[width=\linewidth]{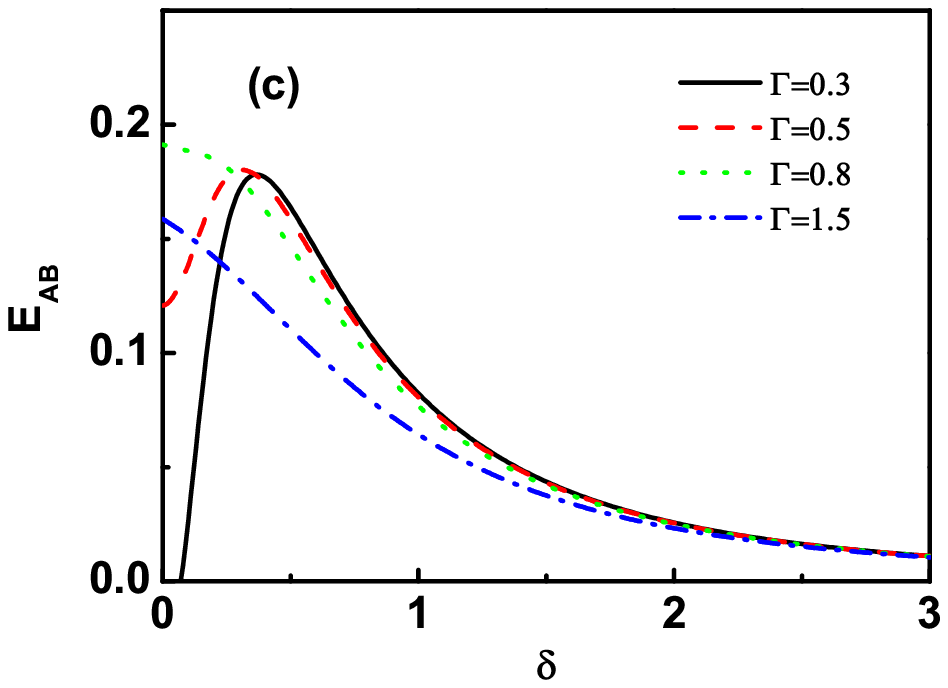}
\caption{(a)and (b)the time evolution of the entanglement for different values of the detuning and $\Gamma$. (c) the steady state  entanglement as quantified by the $E_{AB}$ as a function of the detuning $\delta$ for different $\Gamma$.  }
\end{figure}

\begin{figure}
\includegraphics[width=\linewidth]{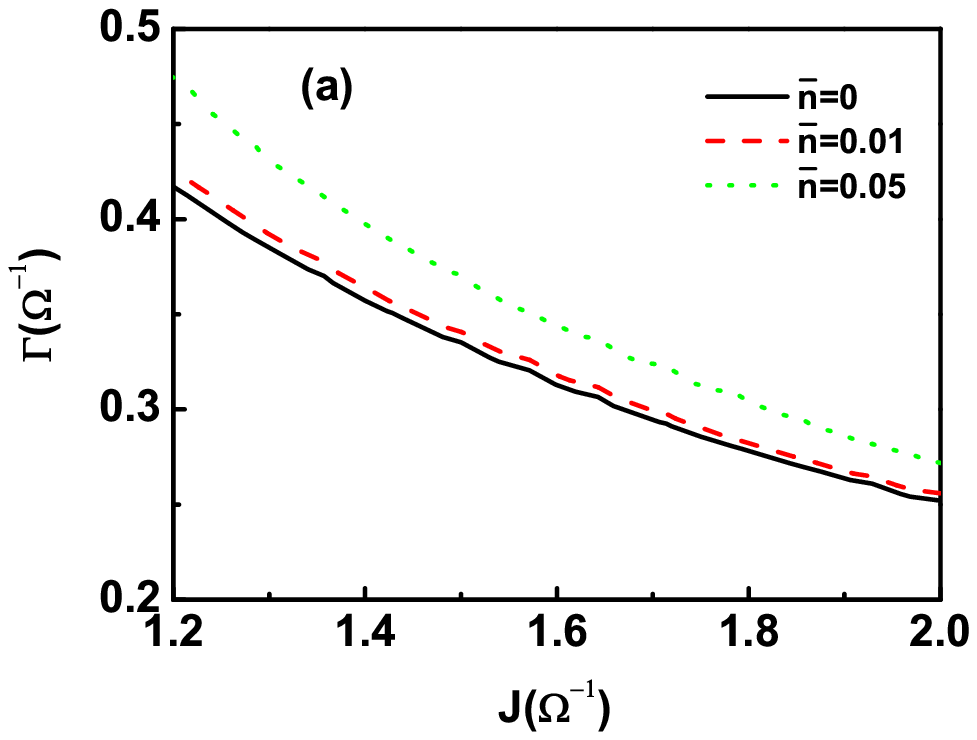}
\includegraphics[width=\linewidth]{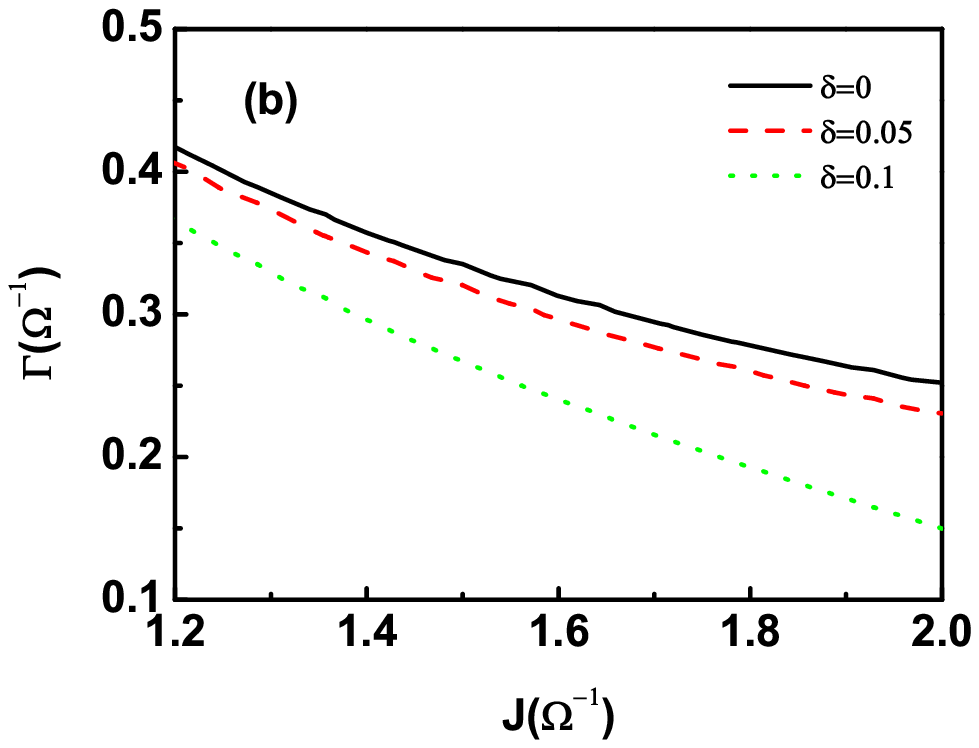}
\caption{The border between presence or absence of entanglement, depending on the  $J$, the noise strength, the detuning  and the bath's
temperature. The parameter (a) $\delta=0$, (a) $\bar{n}=0$.}
\end{figure}

As the reservoir is assumed to be finite-temperature, in Fig.1(c), we can observe the entanglement changes of a slightly different character. The phenomena of ESD and ESB are more evident as the value of $\bar{n}$ is increased.  In the disentanglement dynamics of the finite-temperature dissipation environment, the lifetime, corresponding to nonzero value of entanglement, becomes less, and the exponential disentanglement disappears and ESD or ESB appears. Furthermore, we find the death time and birth time are prolonged and the maximal value that the entanglement can reach decreases with increasing temperature. We also observe steady-state entanglement even for infinite temperature of the bath. The non-monotonicity of quantum entanglement is also apparent, plotted in Fig.1(d) for increasing values of the mean thermal boson number $\bar{n}$. At the given $\Omega$ and $J$, the value of $\Gamma$ that maximizes the steady state entanglement is now a function of $\bar{n}$  and a numerical analysis shows that the critical noise strength $\Gamma_{c}$ of generating entanglement needs higher values as $\bar{n}$ increases. That is to say, strong noise strengths induced the finite-temperature quantum entanglement. However, the peak value of the entanglement tends to decrease faster and the $ \Gamma_{m}$, where quantum entanglement approaches a maximum, shifts to right very rapidly. That is to say, a larger steady-state entanglement can be created for a smaller $\bar{n}$. Such control effect on the entanglement dynamics by varying the bath temperature becomes more significant. At a finite temperature, the steady state entanglement is decreased but remains finite. For an average photon number of $\bar{n}=0.05$, we find that the steady state entanglement reaches the maximal value of 0.1. For GHz frequencies as they are typical for quantum optical implementations this corresponds to around 120 mK. For typical biological systems, however, the bath spectral density peaks around 200 $cm^{-1}$(corresponding to $\omega=4\times 10^{13}$ HZ), so that ($\bar{n}=0.3$)($\bar{n}=0.05$), corresponds to a temperature $T\approx 77$K ($T\approx 30$ K). At these temperatures, typical biomolecular systems such as the Fenna-Matthew-Olson complex exhibits long-lived coherences in the dynamics as demonstrated in recent experiments \cite{Engel} and quantum entanglement survives in photosynthetic light-harvesting complexes despite the decohering effects of their environments\cite{Sarovar,Caruso}. This system could opens up new channels in bath assisted entanglement in a very natural way and even could give more control parameters.

We have considered the the detuning from the qubit and driving frequency is zero, all the phenomena described so far are robust in the presence of a finite $\bar{n}$. Fig.2 shows that the time evolution of the entanglement for different values of the detuning and $\Gamma$. On exact resonance, If $\Gamma<\Gamma_{c}$, the steady state entanglement value $E_{AB}$ for the driven qubits coupled to a bosonic environment at zero and finite temperatures is strictly zero. For a finite detuning, the steady state turns out to be entangled even when the noise strengths below the critical value. The presence of steady state entanglement can be linked unambiguously to an increasing degree of the detuning, the weak noise channel with sufficiently large detuning is an important condition for steady-state entanglement. Quantum correlations increase monotonically up to a maximum corresponding to a certain optimal detuning above which $E_{AB}$ decreases. The reason is that the additional detuning between the qubit and driving frequency produces some new coherences terms. On the contrary, due to the competition of noise strengths and the detuning, if $\Gamma=0.8$,  we can easily see that the value of steady state entanglement becomes smaller with the increase of the detuning and is a monotonically decreasing function of the detuning. This implies that increasing the detuning can not only induce but also suppress steady state entanglement, which depends on the value of noise strength. When the two qubits are subject to system-environment interactions of the same noise strength, the detuning character can be the crucial property that leads to steady state entanglement where purely low noise would result in the complete destruction of entanglement. The above interesting character is well described in Fig.2(c). Anomalous decoherence effects phenomena, as quantified by dynamical quantum entanglement measures, should also be observable in chains of coupled weakly driven spin systems.

From the above analysis, it is clear that the noise strength, the detuning from the qubit frequency  and the finite temperature of independent environments have a notable influence on the steady state entanglement.  The entanglement characteristics shown in Fig.3 suggest that the steady state will fall into either the entangled part or the separable part, i.e. from separable to entangled subsystems, depending on the coupling strength $J$, the noise strength, the detuning and the bath's temperature. The smaller the value for the zero temperature noise, the larger the qubit interaction strength $J$ required for the driven qubits to be entangled. Moreover, we see the points which delimit the border between
presence or absence of entanglement, which is different when varying temperature and the detuning. Alternatively, anomalous decoherence effects can be characterized using a steady entanglement measure of the system's response to the external driving. The steady state of the system can
be computed analytically for $\bar{n}=0, \delta=0$, The system is entangled, and have a negative partial transpose, only if $\Gamma>\Gamma_{c}$, where $\Gamma_{c} =\Omega^{2}/2 J $ is the noise threshold, which is consistent with the numerical results. For the other case, we only give the numerical results. An interesting question is that for the given $\Omega$ and $J$, what is the relations between the maximal steady state entanglement and the value of $\Gamma_{m}$. From Fig.4,  we can find that if the bath's temperature is low enough ($\bar{n}\leq 0.05$), the maximal steady state entanglement is a linearly decreasing function of the critical value of $\Gamma_{m}$. Of course, this relation is not established for the high temperature, and this needs further study.

\begin{figure}
\includegraphics[width=\linewidth]{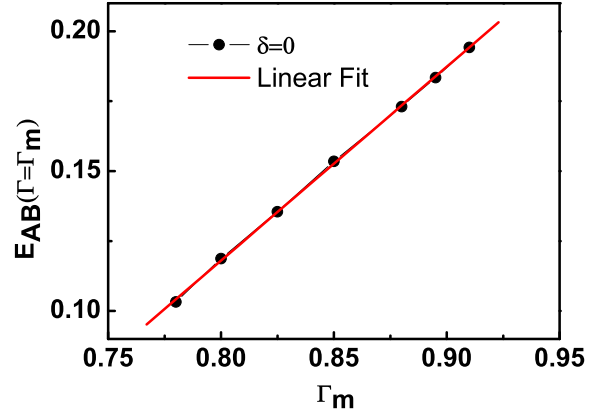}
\caption{The relations between the maximal steady state entanglement and the value of $\Gamma_{m}$ at different bath's
temperature.}
\end{figure}

\section{conclusions}

In conclusion, protecting the dynamics of coupled quantum systems from decoherence by the environment is a key challenge for solid-state quantum information processing. We have studied quantum entanglement dynamics in driven coupled quantum spin systems at zero and finite temperature. When the rotating wave and Born-Markov approximation are used, we reveal numerically that the external driving and  noise strength cause the anomalous decoherence phenomenon, i.e.,  the steady state response of the system,  will be optimized at intermediate noise levels and therefore, trying to reduce the environmental noise to as small as possible values, does not necessarily provide an optimal universal strategy to maximize coherent effects. One actively exploits the noise decay to drive the system to a entangled stationary state which does not depend on the initial states. Classical driving field can be used to stabilize entanglement in this systems. Our results also suggest a control way to beat the effect of decoherence by engineering the noise strength of the reservoirs to approach the steady state entanglement. This demonstrates the possibility of stable steady state entanglement in natural systems consisting of many qubits. We expect these studies to contribute towards the identification of the physical mechanisms that could induce stationary quantum correlations in very noisy environments occurring in experimental conditions.
\section*{Appendix: The solution of quantum master equation }
 In the presence of a classical driving field, the entanglement evolution no longer belongs to the so-called $X$-class state, the density matrix in driven coupled quantum spin systems is of the following form
\begin{align} \rho=\left(
\begin{array}{cccc}
\rho _{11} & \rho _{12} & \rho _{13} & \rho _{14} \\
\rho _{21} & \rho _{22} & \rho _{23} & \rho _{24} \\
\rho _{31} & \rho _{32} & \rho _{33} & \rho _{34}  \\
\rho _{41} & \rho _{42} & \rho _{43} & \rho _{44}
\end{array}\right)\tag{A1}
\end{align}
in the  two-qubit product state basis of $\{\left\vert \uparrow \uparrow \right\rangle ,\left\vert \uparrow \downarrow \right\rangle ,\left\vert \downarrow \uparrow \right\rangle,\left\vert \downarrow \downarrow \right\rangle \}$. Substituting (A1) into (3), i.e. the master equation of our system, we obtain the following first-order coupled differential equations:
\begin{eqnarray}
\frac{d \rho_{11}(t)}{dt}&=&-4(1+\bar{n})\Gamma\rho_{11}(t)+2\bar{n}\Gamma(\rho_{22}(t)+\rho_{33}(t))\nonumber\\
 & &-i\Omega(-\rho_{12}(t)-\rho_{13}(t)+\rho_{21}(t)+\rho_{31}(t))\nonumber\\
\frac{d \rho_{12}(t)}{dt}&=&-3(1+\bar{n})\Gamma\rho_{12}(t)+\bar{n}\Gamma(-\rho_{12}(t)+2\rho_{34}(t))\nonumber\\
 & &-i(-\Omega\rho_{11}(t)+2(\delta-J)\rho_{12}(t)-\Omega\rho_{14}(t)\nonumber\\
 & &+\Omega\rho_{22}(t)+\Omega\rho_{32}(t))\nonumber\\
\frac{d \rho_{13}(t)}{dt}&=&-3(1+\bar{n})\Gamma\rho_{13}(t)+\bar{n}\Gamma(-\rho_{13}(t)+2\rho_{24}(t))\nonumber\\
 & &-i(-\Omega\rho_{11}(t)+2(\delta-J)\rho_{13}(t)-\Omega\rho_{14}(t)\nonumber\\
 & &+\Omega\rho_{23}(t)+\Omega\rho_{33}(t))\nonumber\\
 \frac{d \rho_{14}(t)}{dt}&=&-2(1+2\bar{n})\Gamma \rho_{14}(t)-i(-\Omega \rho_{12}(t)-\Omega \rho_{13}(t)\nonumber\\
 & &+4\delta \rho_{14}(t)+\Omega \rho_{24}(t)+\Omega \rho_{34}(t))\nonumber\\
\frac{d \rho_{21}(t)}{dt}&=&-3(1+\bar{n})\Gamma\rho_{21}(t)+\bar{n}\Gamma(-\rho_{21}(t)+2\rho_{43}(t))\nonumber\\
 & &-i(\Omega\rho_{11}(t)-2(\delta-J)\rho_{21}(t)-\Omega\rho_{22}(t)\nonumber\\
 & &-\Omega\rho_{23}(t)+\Omega\rho_{41}(t))\nonumber\\
\frac{d \rho_{22}(t)}{dt}&=&2(1+\bar{n})\Gamma(\rho_{11}(t)-\rho_{22}(t))-2\bar{n}\Gamma(\rho_{22}(t)\nonumber\\
 & &-\rho_{44}(t))-i\Omega(\rho_{12}(t)-\rho_{21}(t)-\rho_{24}(t)+\rho_{42}(t))\nonumber\\
\frac{d \rho_{23}(t)}{dt}&=&-2(1+2\bar{n})\Gamma\rho_{23}(t)-i\Omega(\rho_{13}(t)-\rho_{21}(t)\nonumber\\
 & &-\rho_{24}(t)+\rho_{43}(t))\nonumber\\
\frac{d \rho_{24}(t)}{dt}&=&(1+\bar{n})(2\rho_{13}(t)-\rho_{24}(t))-3\bar{n}\Gamma\Omega\rho_{24}(t)\nonumber\\
 & &-i(\Omega\rho_{14}(t)-\Omega\rho_{22}(t)-\Omega\rho_{23}(t)+\Omega\rho_{44}(t))\nonumber\\
 & &+2(\delta+J)\rho_{24}(t)\nonumber\\
\frac{d \rho_{31}(t)}{dt}&=&-3(1+\bar{n})\Gamma\rho_{31}(t)+\bar{n}\Gamma(-\rho_{31}(t)+2\rho_{42}(t))\nonumber\\
 & &-i(\Omega\rho_{11}(t)-2(\delta-J)\rho_{31}(t)-\Omega\rho_{32}(t)\nonumber\\
 & &-\Omega\rho_{33}(t)+\Omega\rho_{41}(t))\nonumber\\
\frac{d \rho_{32}(t)}{dt}&=&-2(1+2\bar{n})\Gamma\rho_{32}(t)-i\Omega(\rho_{12}(t)-\rho_{31}(t)\nonumber\\
 & &-\rho_{34}(t)+\rho_{42}(t))\nonumber\\
\frac{d \rho_{33}(t)}{dt}&=& 2(1+\bar{n})\Gamma(\rho_{11}(t)-\rho_{33}(t))-2\bar{n}\Gamma(\rho_{33}(t)\nonumber\\
 & &-\rho_{44}(t))-i\Omega(\rho_{13}(t)-\rho_{31}(t)-\rho_{34}(t)+\rho_{43}(t))\nonumber\\
\frac{d \rho_{34}(t)}{dt}&=&(1+\bar{n})(2\rho_{12}(t)-\rho_{34}(t))-3\bar{n}\Gamma\Omega\rho_{34}(t)\nonumber\\
 & &-i(\Omega\rho_{14}(t)-\Omega\rho_{32}(t)-\Omega\rho_{33}(t)+\Omega\rho_{44}(t)\nonumber\\
 & &+2(\delta+J)\rho_{34}(t))\nonumber\\
\frac{d \rho_{41}(t)}{dt}&=&-2(1+2\bar{n})\Gamma\rho_{41}(t)-i(\Omega \rho_{21}(t)+\Omega \rho_{31}(t)\nonumber\\
 & &-4\delta \rho_{41}(t)-\Omega \rho_{42}(t)-\Omega \rho_{43}(t))\nonumber\\
\frac{d \rho_{42}(t)}{dt}&=&(1+\bar{n})(2\rho_{31}(t)-\rho_{42}(t))-3\bar{n}\Gamma\rho_{42}(t)\nonumber\\
 & &-i(\Omega\rho_{22}(t)+\Omega\rho_{32}(t)-\Omega\rho_{41}(t)-\Omega\rho_{44}(t)\nonumber\\
 & &-2(\delta+J)\rho_{42}(t))\nonumber\\
\frac{d \rho_{43}(t)}{dt}&=&(1+\bar{n})(2\rho_{21}(t)-\rho_{43}(t))-3\bar{n}\Gamma\rho_{43}(t)\nonumber\\
 & &-i(\Omega\rho_{23}(t)+\Omega\rho_{33}(t)-\Omega\rho_{41}(t)-\Omega\rho_{44}(t)\nonumber\\
& &-2(\delta+J)\rho_{43}(t))\nonumber\\
\frac{d \rho_{44}(t)}{dt}&=&2(1+\bar{n})(\rho_{22}(t)+\rho_{33}(t))-4\bar{n}\Gamma\rho_{44}(t)\nonumber\\
& &-i\Omega(\rho_{24}(t)+\rho_{34}(t)-\rho_{42}(t)-\rho_{43}(t))\nonumber\\
\end{eqnarray}

Obviously, the solution of (A2) depends on the initial state of the qubits, so we can solve analytically  and numerically for some typical initial states. By solving the equations of $\frac{d \rho}{dt}=0$ to get the steady-state solutions, and then we can study the
steady-state properties of the two driven coupled qubits.


\begin{thebibliography}{00}

\bibitem{Nielsen}
M. A. Nielsen, I. L. Chuang, Quantum Computation and Quantum Information (Cambridge University Press, Cambridge, 2000).

\bibitem{Yu}
T. Yu, J. H. Eberly, Finite-Time Disentanglement Via Spontaneous Emission, Phys. Rev. Lett. 93, 140404 (2004);
T. Yu and J. H. Eberly, Entanglement sudden death, Science 323, 598 (2009).

\bibitem{experimental}
L. Aolita, R. Chares, D. Cavalcanti, A. Acln, and L. Davidovich, Phys. Rev. Lett. 100, 080501 (2008); M. P. Almeida, F. de Melo, M. Hor-Meyll, A. Salles, S. P. Walborn, P. H. Souto Ribeiro, and L. Davidovich, Science 316, 579 (2007).

\bibitem{Ficek}
Z. Ficek and R. Tanas, Delayed sudden birth of entanglement, Phys. Rev. A 77, 054301 (2008).
\bibitem{Lopez}
C. E. Lopez, G. Romero, F. Lastra, E. Solano, and J. C. Retamal, Sudden birth versus sudden death of entanglement in multipartite
systems, Phys. Rev. Lett. 101, 080503 (2008).


\bibitem{Xu}
J. S. Xu, X. Y. Xu, C. F. Li, C. J. Zhang, X. B. Zou, and G. C. Guo, Experimental investigation of classical and quantum correlations
under decoherence,  Nat. Commun. 1, 1 (2010).
\bibitem{Mazzola}
L. Mazzola, S. Maniscalco, J. Piilo, K. A. Suominen, and B. M. Garraway, Sudden death and sudden birth of entanglement in common structured reservoirs, Phys. Rev. A 79, 042302 (2009).
\bibitem{Dijkstra}
A. G. Dijkstra and Y. Tanimura, Non-Markovian Entanglement Dynamics in the Presence of System-Bath Coherence, Phys. Rev. Lett. 104, 250401 (2010).
\bibitem{Jing}
J. Jing, L. -A. Wu, Marcelo. S. Sarandy, and J. Gonzalo, Muga, Inverse engineering control in open quantum systems, Phys. Rev. A 88, 053422 (2013).
\bibitem{Shin}
C. S. Shin, C. E. Avalos, M. C. Butler, H. J. Wang, S. J. Seltzer, R. -B. Liu, A. Pines, V. S. Bajaj, Suppression of electron spin decoherence of the diamond NV center by a transverse magnetic field, Phys. Rev. B 88, 161412(R) (2013).


\bibitem{Huelga}
S. F. Huelga, A. Rivas, and M. B. Plenio, Non-Markovianity-Assisted Steady State Entanglement, Phys. Rev. Lett. 108, 160402 (2012).
\bibitem{Verstraete}
F. Verstraete, M. M. Wolf, and J. I. Cirac, Quantum computation and quantum-state engineering driven by dissipation, Nature Phys. 5, 633 (2009).
\bibitem{Kraus}
B. Kraus, H. P. Buchler, S. Diehl, A. Kantian, A. Micheli, and P. Zoller, Preparation of entangled states by quantum Markov processes,
Phys. Rev. A 78, 042307 (2008).
\bibitem{Plenio}
 M. B. Plenio and S. F. Huelga, Entangled Light from White Noise, Phys. Rev. Lett. 88, 197901 (2002).
\bibitem{Wolf}
A. Wolf, G. De Chiara, E. Kajari, E. Lutz, and G. Morigi, Entangling two distant oscillators with a quantum reservoir, Europhys. Lett. 95, 60008 (2011).
\bibitem{Bellomo}
B. Bellomo, R. L. Franco, and G. Compagno, Non-Markovian Effects on the Dynamics of Entanglement, Phys. Rev. Lett. 99, 160502 (2007).
B. Bellomo, R. L. Franco, S. Maniscalco, and G. Compagno, Entanglement trapping in structured environments, Phys. Rev. A 78, 060302(R) (2008).
\bibitem{Hartmann}
L. Hartmann, W. Dur, and H. J. Briegel, Steady-state entanglement in open and noisy quantum systems, Phys. Rev. A 74, 052304 (2006).
L. Hartmann, W. Dur, and H. J. Briegel, Entanglement and its dynamics in open, dissipative systems, New J. Phys. 9, 230 (2007).

\bibitem{Liu}
N. Zhao, Z. Y. Wang, and R. B. Liu, Anomalous decoherence effect in a quantum bath, Phys. Rev. Lett. 106, 217205 (2011).
\bibitem{An}
H. B. Liu, J. H. An, C. Chen, Q. J. Tong, H. G. Luo, and C. H. Oh, Anomalous decoherence in a dissipative two-level system,
 Phys. Rev. A 87, 052139 (2013).
\bibitem{Du}
P. Huang, X. Kong, N. Zhao, F. Shi, P. Wang, X. Rong, R. B. Liu, J. Du, Observation of an anomalous decoherence effect in a quantum bath at room temperature, Nat. Commun. 2, 570 (2011).

\bibitem{Breuer}
 H. P. Breuer and F. Petruccione, The Theory of Open Quantum Systems (Oxford University Press, New York, 2002).

\bibitem{Poyatos}
J. F. Poyatos, J. I. Cirac, and P. Zoller, Quantum Reservoir Engineering with Laser Cooled Trapped Ions, Phys. Rev. Lett. 77, 4728 (1996).
\bibitem{Diehl}
S. Diehl,  A. Micheli, A. Kantian, B. Kraus, H. P. B¨¹chler, and P. Zoller, Quantum states and phases in driven open quantum systems with cold atoms,
Nature Phys. 4, 878 (2008).
\bibitem{Prior}
J. Prior, A. W. Chin, S. F. Huelga, and M. B. Plenio, Efficient Simulation of Strong System-Environment Interactions, Phys. Rev. Lett. 105, 050404 (2010).
\bibitem{Wang}
Z. H. Wang, B. S. Wang, Z. B. Su, Entanglement evolution of a spin chain bath in driving the decoherence of a coupled quantum spin, Phys. Rev. B 79, 104428 (2009).

\bibitem{Wu}
L. A. Wu, G. Kurizki, and P. Brumer, Master Equation and Control of an Open Quantum System with Leakage, Phys. Rev. Lett. 102, 080405 (2009).



\bibitem{Yao}
W. Yang, and  R. B. Liu, Quantum many-body theory of qubit decoherence in a finite-size spin bath. Phys. Rev. B 78 , 085315 (2008).
\bibitem{Huelga and Plenio}
S. F. Huelga and M. B. Plenio, Stochastic Resonance Phenomena in Quantum Many-Body Systems, Phys. Rev. Lett. 98, 170601 (2007).
\bibitem{Liao}
J.-Q. Liao, J.-F. Huang, L.-M. Kuang, and C. P. Sun, Coherent excitation-energy transfer and quantum entanglement in a dimer, Phys. Rev. A 82, 052109 (2010).

\bibitem{Lambert}
N. Lambert, R. Aguado, and T. Brandes, Nonequilibrium entanglement and noise in coupled qubits, Phys. Rev. B 75, 045340 (2007).
\bibitem{Li}
J. Li and G. S. Paraoanu, Generation and propagation of entanglement in driven coupled-qubit systems, New J. Phys. 11, 113020 (2009).
\bibitem{Galve}
F. Galve, L. A. Pachon, and D. Zueco, Bringing Entanglement to the High Temperature Limit, Phys. Rev. Lett.
105, 180501 (2010).
\bibitem{Cai}
J. Cai, S. Popescu, and H.-J. Briegel, Dynamic entanglement in oscillating molecules and potential biological implications, Phys. Rev. E 82,
021921 (2010).
J. Cai, G. G. Guerreschi,  and H.-J. Briegel, Quantum Control and Entanglement in a Chemical Compass, Phys. Rev. Lett. 104, 220502 (2010).

\bibitem{Porras}
D. Porras and J. I. Cirac, Effective Quantum Spin Systems with Trapped Ions, Phys. Rev. Lett. 92, 207901 (2004);
K. Kim, M.-S. Chang, S. Korenblit, R. Islam, E. E. Edwards, J. K. Freericks, G.-D. Lin, L.-M. Duan and C. Monroe, Quantum simulation of frustrated Ising spins with trapped ions, Nature 465, 590 (2010);
J. Simon, W. S. Bakr, R. Ma, M. Eric Tai, P. M. Preiss, and M. Greiner, Quantum simulation of antiferromagnetic spin chains in an optical lattice, Nature 472, 307 (2011).


\bibitem{cohen}
C. Cohen-Tannoudji, J. Dupont-Roc and G. Grynberg, Atom-Photon Interactions, (Wiley, New York, 1992).

\bibitem{Vidal}
G. Vidal, R. F. Werner, Computable measure of entanglement, Phys. Rev. A 65, 032314 (2002).
\bibitem{Peres}
A. Peres, Separability Criterion for Density Matrices, Phys. Rev. Lett. 77, 1413 (1996).
\bibitem{Horodecki}
M. Horodecki, P. Horodecki, R. Horodecki, Separability of Mixed States: Necessary and Sufficient Conditions, Physics Letters A 223, 1 (1996).


\bibitem{Engel}
G. S. Engel, T. R. Calhoun, E. L. Read, T. K. Ahn, T. Mancal, Y.-C. Cheng, R. E. Blankenship, and G. R. Fleming, Evidence for wavelike energy transfer through quantum coherence in photosynthetic systems, Nature (London) 446, 782 (2007).
\bibitem{Sarovar}
M. Sarovar, A. Ishizaki, G. R. Fleming, and K. B. Whaley, Quantum entanglement in photosynthetic light-harvesting complexes, Nature Physics 6, 462 (2010).
\bibitem{Caruso}
F. Caruso, A. W. Chin, A. Datta, S. F. Huelga, and M. B. Plenio, Entanglement and entangling power of the dynamics in light-harvesting complexes,
Phys. Rev. A 81, 062346 (2010).


\end{thebibliography}
\end{document}